# A gated group sequential design for seamless Phase II/III trial with subpopulation selection

Guanhong Miao[1], Jason J.Z. Liao[2*], Jing Yang[3], Keaven Anderson[3]

[1]Department of Biostatistics, University of Florida, Gainesville, FL 32611
[2]Biostatistics, Incyte Corporation, Willington, DE 19803; jliao@incyte.com
[3]BARDS, Merck & Co., Inc., North Wales, PA 19454

**Abstract**

Due to the high cost and high failure rate of Phase III trials, seamless Phase II/III designs are more and more popular to trial efficiency. A potential attraction of Phase II/III design is to allow a randomized proof-of-concept stage prior to committing to the full cost of the Phase III trial. Population selection during the trial allows a trial to adapt and focus investment where it is most likely to provide patient benefit. Motivated by a clinical trial to find the population that potential benefits with dual-primary endpoints progression free survival (PFS) and overall survival (OS), we propose a gated group sequential design for a seamless Phase II/III trial design with population selection. The investigated design controls the familywise error rate and allows multiple interim analyses to enable early stopping for efficacy or futility. Simulations and an illustrative example suggest that the proposed gated group sequential design can have more power than the commonly used classical group sequential design, and reduces the patient's exposure to less effective treatment if the complementary sub-group has less significant treatment effect. The proposed design has the potential to save drug development cost and more quickly fulfill unmet medical needs.

## 1. Introduction

The high failure rate of phase III trials combined with their substantial cost make selecting an appropriate treatment and population for evaluation of paramount importance (Pretorius and Grignolo, 2016). Seamless Phase II/III multi-arm clinical trials, which use the initial part of the trial (Phase II) to learn about all treatments while an in-depth evaluation occurs only on the promising one(s) in the second part (Phase III), are potentially efficient to overcome this challenge. Using data accumulated across both phases of a single Phase II/III trial may enable more efficient development of a treatment for an appropriate population than separate trials for Phases II and III.

Consider planning for a second line small cell lung cancer clinical trial, where the literature suggests a platinum-sensitive sub-group may yield a much greater treatment benefit. Even if the treatment benefit in the platinum-resistant sub-group is less certain, from a marketing perspective, the all-comer population with the inclusion of the platinum-resistant sub-group can give maximum patient benefit, followed by market value if the platinum-resistant sub-group also receives benefit from the experimental treatment. Under this circumstance, a direct Phase III trial with a broad population can be risky. A more efficient approach could be a seamless Phase II/III design with population selection in the Phase II portion of the trial followed by a potentially targeted Phase III enrollment with focused patient population to confirm the benefit.

In clinical trials, the clinical benefit of an intervention is often characterized by multiple outcomes. For multiple hypothesis testing problems, the familywise error rate (FWER), i.e., the probability to erroneously reject at least one null hypothesis, needs to be controlled and bounded by a pre-specified significance level α. A sequence of methods derived from weighted Bonferroni-based closed test procedures have been proposed to control the FWER for multiple testing. Examples of such methods include Bonferroni-Holm procedure [Holm, 1979], gatekeeping procedures based on Bonferroni adjustments [Chen et al., 2005] and the graphical approach [Edwards and Madsen, 2007; Maurer and Bretz, 2013]. As group sequential designs are widely used and commonly employed in order to facilitate early efficacy testing, the application of group sequential designs to multiple endpoints becomes popular and has been widely studied recently [Hung et al., 2007; Liu and Anderson, 2008; Glimm et al., 2010; Tamhane et al., 2010, 2012a &b; Maurer and Bretz, 2013; Asakura et al., 2014; Hamasaki et al., 2015; Schuler et al., 2017; Xu et al., 2018].

Adaptive seamless Phase II/III designs allow Phase II assessment of whether within-trial extension to Phase III is justified. Here we consider that the adaptation includes choosing a meaningful population for an effective investment with high probability of success. A pre-defined, targeted sub-group and the full population are both studied in the first stage of the adaptive Phase II/III design. Investment in the second stage of the adaptive Phase II/III design is then focused on the population(s) that are most likely to provide patient benefit after the futility analysis at the end of Phase II. Due to the multiple sources potentially contributing to the decision error in this type of design, the FWER control should be studied carefully. The closed testing procedure [Marcus et al., 1976] is usually applied to test multiple hypotheses in the setting of population selection. The FWER control strategies using multiple testing method [Simes, 1986; Spiessens and Debois, 2010], combination test method [Bretz et al., 2006; Brannath et al., 2009], the marginal p-value combinational approach (Sugitani, Bretz and Maurer, 2016), and a conditional error function approach [Scala and Glimm, 2011] have been proposed. If the interest is in multiple outcomes, the application of adaptive Phase II/III designs to multiple endpoints has been investigated using different analysis methods [Stallard, 2010; Jenkins et al., 2011; Friede et al., 2011, 2012].

Group sequential design (GSD) which can terminate the trial early at the interim analysis with sufficient evidence can be combined with the adaptive Phase II/III design. This enables the population selection and treatment benefit confirmation within a single trial. However, the closed testing principle between the sub-group and the full population could dramatically decrease the power of an adaptive Phase II/III design when only one group has meaningful efficacy. Inspired by Glimm et. al. (2010), the hierarchical testing, which was originally used for the testing order of endpoints, can be applied to test the sub-group and full population. In this paper, a gated group sequential design ($g$GSD) is proposed to improve the power by increasing the significance level for each group while controlling FWER. Section 2 gives the details of the proposed design. The performance of *g*GSD is investigated by simulations in section 3, and an illustrative example is used to illustrate the design and its efficiency in section 4. Finally, the summary is given in section 5.

## 2. Methods

We consider a randomized, parallel group clinical trial with two treatment arms -- experimental and control, and dual primary endpoints -- arbitrarily OS and PFS as a prototypical example, thus, a positive study can be claimed as long as one of the dual endpoints achieves its goal. There is an interest to investigate the efficacy of the experimental treatment in both the full population (F) and a targeted sub-group (S). Four null hypotheses below are of interest,

1) $H_0^{\{F,OS\}}$ : no difference in OS between arms in the full population;
2) $H_0^{\{F,PFS\}}$ : no difference in PFS between arms in the full population;
3) $H_0^{\{S,OS\}}$ : no difference in OS between arms in the targeted sub-group;
4) $H_0^{\{S,PFS\}}$ : no difference in PFS between arms in the targeted sub-group.

Let $\alpha_1$, $\alpha_2$, $\alpha_3$ and $\alpha_4$ be the initial alpha level through the graphical approach assigned to test the hypotheses $H_0^{\{F,OS\}}$, $H_0^{\{F,PFS\}}$, $H_0^{\{S,OS\}}$ and $H_0^{\{S,PFS\}}$, respectively, and α be the overall significance level. Jenkins, et al. (2011) proposed a method for population selection in the seamless adaptive design framework with only one analysis in stage 2 after selecting the population in stage 1. In this paper, we extend their method for population selection to control FWER for all four of the aforementioned hypotheses. We further add a group sequential design strategy into stage 2 for flexible early efficacy testing. The design consists of an initial learning stage (stage 1) analogous to a randomized Phase II trial and a second confirmatory phase (stage 2) analogous to a randomized Phase III trial. The selection between populations F and S is based on the PFS results at the end of stage 1. Based on that, the trial can either stop for futility, or continue to stage 2 in both populations F and S, or the sub-group S only, or the full population F only without analyzing the sub-group S in stage 2. In stage 2, we consider group sequential setting with $K-1$ interim analyses and one final analysis, where PFS and OS in populations F and/or S are tested by using group sequential approaches, and alpha allocation follows the graphical approach (Maurer and Bretz, 2013). Figure 1 shows the analysis flowchart for K=3.

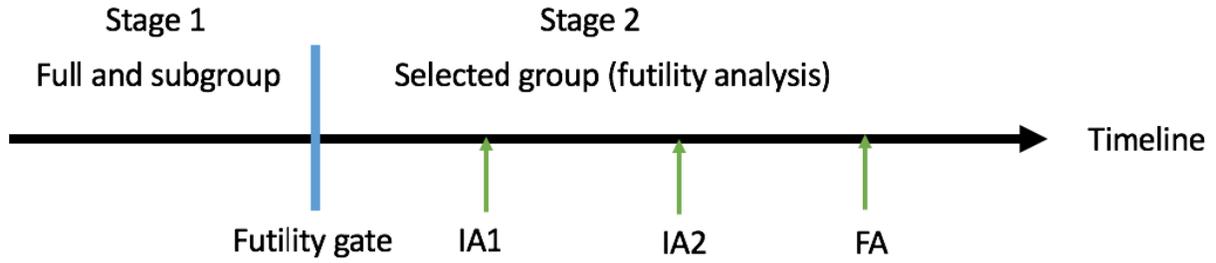

**Figure 1**: Analysis flowchart for the seamless Phase II/III design. IA: interim analysis; FA: final analysis.

According to the FDA guidance for adaptive design (2019), the design, conduct, and analysis of an adaptive clinical trial intended to provide substantial evidence of effectiveness should satisfy four key principles: 1) the chance of erroneous conclusions should be adequately controlled, 2) estimation of treatment effects should be sufficiently reliable, 3) details of the design should be completely pre-specified, and 4) trial integrity should be appropriately maintained. There are three sources for potential inflation of Type I error: 1) early rejection of null hypothesis at interim analysis; 2) adaptation of design features and combination of information across trial stages; and 3) multiple hypothesis testing. To control the type I error rate, the following strategies are proposed: group sequential plans for early rejection; the combinations of p-values using methods such as the inverse normal method for adaptation; the multiple testing methodologies such as the closed testing procedures for multiple hypothesis. If needed, all three approaches can be combined to control the FWER.

For subjects recruited in stage 1, the nominal one-sided observed p-values of $H_0^{\{F,\cdot\}}$ and $H_0^{\{S,\cdot\}}$ at the $k$th analysis ($k = 1,2,\ldots,K$) will be denoted by $p_{1k}^{\{F,\cdot\}}$ and $p_{1k}^{\{S,\cdot\}}$, respectively. For subjects recruited in stage 2, the nominal one-sided observed p-values of $H_0^{\{F,\cdot\}}$ and $H_0^{\{S,\cdot\}}$ at the $k$th analysis ($k = 1,2,\ldots,K$) will be denoted by $p_{2k}^{\{F,\cdot\}}$ and $p_{2k}^{\{S,\cdot\}}$. The goal is to control the FWER at a nominal level α, that is, the probability of rejecting at least one of the true null hypotheses $H_0^{\{F,OS\}}$, $H_0^{\{S,OS\}}$, $H_0^{\{F,PFS\}}$ and $H_0^{\{S,PFS\}}$. We consider all potential sources of error inflation for type I error control, with the closed testing principle applied for multiple testing, inverse combination testing used to analyze the data from two stages, and graphical approach applied for group sequential with different endpoints. Combining these strategies, the FWER of the proposed design is strictly controlled. (Magirr, et. al., 2016; Burnett and Jennison, 2021)

<u>At the end of stage 1: the futility analysis</u>

The futility analysis for PFS in the sub-group $S$ and full population $F$ are performed at the end of stage 1. This determines whether the trial can continue to stage 2 with one or two populations, or just stop at the end of stage 1. Let $HR^F$ and $HR^S$ be the estimated hazard ratio (HR) of the full population and the sub-group, and $\theta^F$ and $\theta^S$ be the pre-specified hazard ratio threshold for the full population and the sub-group, respectively. Table 1 provides the decision rule for population selection. We choose $\theta^x$ ($x = F, S$) to ensure that $P(HR > \theta^x | true\ HR) = \gamma^x$ where $\gamma^x$ is a pre-specified threshold that the trial does not pass the futility gate under the true alternative HR. Under equal randomization, log(HR) approximately follows a normal distribution with mean log(true HR) and variance 4/(number of events). This gives a way to calculate the aforementioned thresholds.

**Table 1**: Population selection rule at the end of stage 1

|  | $HR^F < \theta^F$ | $HR^F \geq \theta^F$ |
|---|---|---|
| $HR^S < \theta^S$ | continue for F and S | continue for S only |
| $HR^S \geq \theta^S$ | continue for F only | stop for futility |

Stage 2:

Once the futility boundary at the end of stage 1 is passed, the trial will continue to stage 2 with one or two populations. As described above, there are three possible scenarios in stage 2.
***Scenario 1****: continue to stage 2 in the sub-group S only with the planned sample size in S, allocating additional alpha to S,* i.e., $\alpha_1 = \alpha_2 = 0$;
***Scenario 2****: continue to stage 2 in the full population F with the planned sample size in F without further analysis of S, allowing additional allocation of alpha to F,* i.e., $\alpha_3 = \alpha_4 = 0$;
***Scenario 3****: continue to stage 2 in both populations F and S with the planned sample size, continuing testing in both populations.*
In this paper, we propose a gated group sequential design (*g*GSD*)* by incorporating the hierarchical testing strategy to the group sequential design. The hierarchical testing strategy was proposed by Glimm et. al. (2010) for the ordered testing of endpoints such as PFS and OS with FWER controlled. However, we modify their strategy to accommodate our multiple testing scenarios with FWER controlled between populations; i.e., the hierarchical testing strategy is used for the ordered testing of populations. Specifically, the approach for stage 2 is as follows after combination of phase 2 and 3 data for testing in Phase 3.

In *scenario 1*, only PFS and OS in the subgroup *S* will be tested according to the alpha allocated using the graphical approach. However, a special graphical approach can also be used, i.e., $H_0^{\{S,PFS\}}$ is first tested with a significance level α, and the full α will be passed to test $H_0^{\{S,OS\}}$ when $H_0^{\{S,PFS\}}$ is rejected. Note that the patients for the F minus S population enrolled in stage 1 will be followed continuously since the information from those patients is needed in the closed testing procedure.
In *scenario 2,* only PFS and OS in the full population *F* will be tested according to the alpha allocated using the graphical approach; analogous to *Scenario 1, a special graphical approach can also be used; i.e.,* $H_0^{\{F,OS\}}$ will be tested at level α only if $H_0^{\{F,PFS\}}$ is rejected.
In *scenario 3,* the sub-group *S* and the full population *F* are tested hierarchically, i.e., the hypotheses in *F* will not be tested until at least one hypothesis in *S* is rejected. For the hypotheses within the same population *F* or *S*, the graphical approach of Maurer and Bretz (2013) is applied. More specifically, the hypotheses in the sub-group *S* can be tested based on the graphical approach with $\alpha_3 + \alpha_4 = \alpha$. Under the hierarchical rule, the hypotheses in the full population $F$ will be tested by using graphical approach with $\alpha_1 + \alpha_2 = \alpha$ when at least one hypothesis in the sub-group $S$ has been rejected. The graphical approach ensures that α-reallocation occurs only between PFS and OS within the same group, and does not occur between different groups (i.e., between $F$ and $S$). Note that the sequential testing rules and the timing of analyses can be totally independent between the sub-group and the full population. Figure 2 below illustrates the *g*GSD testing procedures in stage 2 for the efficacy analyses with *K=3*.

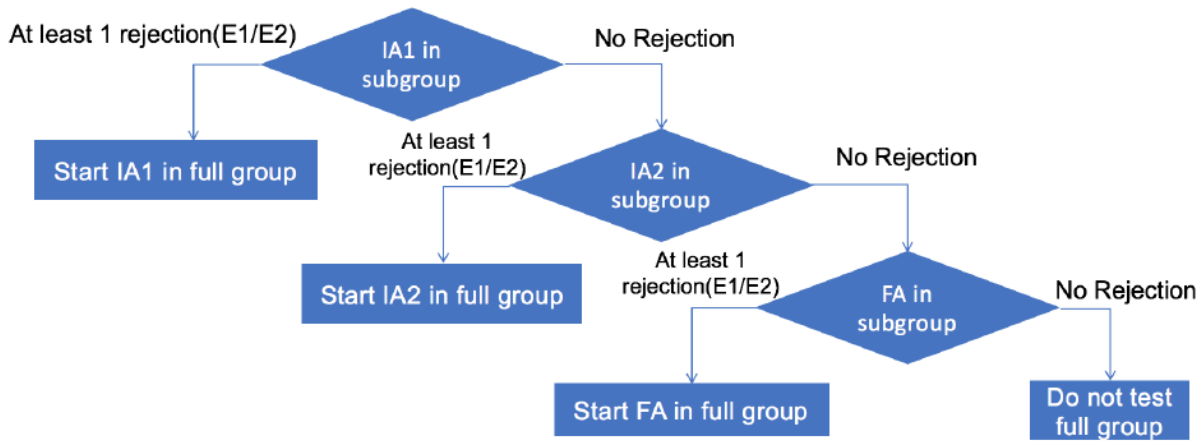

**Figure 2**: The testing procedures based on the gated group sequential design with K=3. E: endpoint; IA: interim analysis; FA: final analysis.

3. **Simulations**

To illustrate the performance of the proposed design in terms of type I error and power, we conduct simulations and compare the performance with the other two well-established approaches:
- GSD: group sequential design for the 4 hypotheses of interest using the graphical approach of Maurer and Bretz (2013) without any population or hypothesis adaptation.
- AD: this is similar to the proposed *g*GSD with subpopulation selection in the futility analysis except that the overall significance level is set to be $\alpha$ to test all 4 hypotheses rather than setting the overall significance level to be $\alpha$ to test only 2 hypotheses in each population (S and F) in gGSD. The same alpha reallocation strategy, graphical approach of Maurer and Bretz (2013) is used to control the FWER.

AD and *g*GSD are applied to the trial with two stages, and GSD is a single-stage design with no adaptation of hypotheses tested at the end of an initial stage. Three simulation settings are considered. In each setting, two interim analyses and one final analysis are planned in stage 2.

Specifically, PFS testing is planned at IA1 and IA2 (which is also the final for PFS), while OS testing is planned at IA1, IA2 and FA. Some parameters are common for all three settings: 1) for the control arm, the median PFS (OS) is assumed to be 4 (10.5) months and 3 (5.7) months both in the sub-group and the complement of the sub-group, respectively; 2) the yearly dropout rates for PFS and OS are 10% and 1%, respectively. In settings 1 and 2, the hazard ratio (experimental/control) for PFS and OS are 0.7 for both the sub-group and the full population. In setting 3, the hazard ratios of PFS and OS are 0.7 for the sub-group, but 1 for the full population. For the full population: at the design stage, the fractions for PFS are approximately 90% for IA1 and IA2 is the final analysis; the fractions for OS are approximately 69% for IA1 and 92% for IA2. For the sub-group population: at the design stage, the fractions for PFS are approximately 89% for IA1 and IA2 is the final analysis; the fractions for OS are approximately 66% for IA1 and 91% for IA2. Some other parameters used in the simulations are provided in Table 2 below where the sample size is calculated based on the group sequential design with a power of 85%. The alpha boundaries are computed using the Lan-DeMets spending function approximating O'Brien-Fleming bounds with a total of 1-sided alpha 2.5%.

**Table 2**: Parameters for three simulation scenarios

| Setting | Sample Size | Subgroup Proportion | Enrollment Duration | Design | Full Group | | Sub-Group | |
|---|---|---|---|---|---|---|---|---|
| | | | | | $\alpha_1$ (OS) | $\alpha_2$ (PFS) | $\alpha_3$ (OS) | $\alpha_4$ (PFS) |
| 1 | 554 | 0.75 | 28 | GSD & AD | 0.22% | 0.165% | 1.28% | 0.835% |
| | | | | $g$GSD | 1.429% | 1.071% | 1.513% | 0.987% |
| 2 & 3 | 924 | 0.5 | 33 | GSD & AD | 0.025% | 0.017% | 1.458% | 1.00% |
| | | | | $g$GSD | 1.488% | 1.012% | 1.48% | 1.02% |

For each setting, the performance of GSD is provided as a reference for comparison. For AD and $g$GSD, the futility analyses for PFS are performed at the end of stage 1. This determines whether the trial continues to stage 2 with one or two populations, or the trial stops. Let the futility threshold, $\gamma$, the probability of the trial not passing the futility gate under the alternate hypothesis, be 5%. This results in $\theta^F = 0.85$ and $\theta^S = 0.9$ for setting 1, and $\theta^F = 0.83$ and $\theta^S = 0.85$ for settings 2 and 3.

The inverse-normal combination test is applied to control the FWER regardless of the decision at the futility analysis at the end of stage 1. For the *k-th* analysis in stage 2, weights $w_{1k}$ and $w_{2k}$ are pre-specified to combine the p-values from stage 1 ($p_{1k}$) and stage 2 ($p_{2k}$), where $w_{1k}^2 + w_{2k}^2 = 1$. The null hypothesis is rejected if $w_{1k}\Phi^{-1}(1 - p_{1k}) + w_{2k}\Phi^{-1}(1 - p_{2k}) \geq c_k$, where $c_k$ is the z-statistic boundary using the allocated alpha. Closed testing procedures are applied to control the FWER as we apply the population selection in the design. The Hochberg correction [Hochberg, 1988] with equal weighting, $p_i^{FS} = \min[2\min\{p_i^F, p_i^S\}, max\{p_i^F, p_i^S\}]$, is used to compute the p-values of the intersection hypotheses between the populations. The minimum z-statistic boundary of hypotheses with the allocated alpha in the intersection testing is used as $c_k$.

The weights and p-values to be used in combination tests are provided below, where the PFS endpoint is used as an example; the OS endpoint can be performed in a similar manner. Note that the weights $w_{1k}$ and $w_{2k}$ need to be pre-specified for controlling the FWER, and can be different for PFS and OS endpoints.

1. S only scenario – when considering $H_0^{\{S,PFS\}}$ only.
   Testing $H_0^{\{FS,PFS\}}$: $w_{1k}\Phi^{-1}\left(1-p_{1k}^{\{FS,PFS\}}\right) + w_{2k}\Phi^{-1}\left(1-p_{2k}^{\{S,PFS\}}\right)$;
   Testing $H_0^{\{S,PFS\}}$: $w_{1k}\Phi^{-1}\left(1-p_{1k}^{\{S,PFS\}}\right) + w_{2k}\Phi^{-1}\left(1-p_{2k}^{\{S,PFS\}}\right)$.

2. F only scenario - when considering $H_0^{\{F,PFS\}}$ only.
   Testing $H_0^{\{FS,PFS\}}$: $w_{1k}\Phi^{-1}\left(1-p_{1k}^{\{FS,PFS\}}\right) + w_{2k}\Phi^{-1}\left(1-p_{2k}^{\{F,PFS\}}\right)$;
   Testing $H_0^{\{F,PFS\}}$: $w_{1k}\Phi^{-1}\left(1-p_{1k}^{\{F,PFS\}}\right) + w_{2k}\Phi^{-1}\left(1-p_{2k}^{\{F,PFS\}}\right)$.

3. F and S scenario – when considering both $H_0^{\{F,PFS\}}$ and $H_0^{\{S,PFS\}}$.
   Testing $H_0^{\{FS,PFS\}}$: $w_{1k}\Phi^{-1}\left(1-p_{1k}^{\{FS,PFS\}}\right) + w_{2k}\Phi^{-1}\left(1-p_{2k}^{\{FS,PFS\}}\right)$;
   Testing $H_0^{\{F,PFS\}}$: $w_{1k}\Phi^{-1}\left(1-p_{1k}^{\{F,PFS\}}\right) + w_{2k}\Phi^{-1}\left(1-p_{2k}^{\{F,PFS\}}\right)$;
   Testing $H_0^{\{S,PFS\}}$: $w_{1k}\Phi^{-1}\left(1-p_{1k}^{\{S,PFS\}}\right) + w_{2k}\Phi^{-1}\left(1-p_{2k}^{\{S,PFS\}}\right)$.

A total of 2000 replications are performed for each setting. For AD and $g$GSD, eight different sets of weights were evaluated for the inverse-normal combination tests. Ideally, weights $w_{1k}$ and $w_{2k}$ would be chosen to be proportional to the square root of the number of events in each stage for the $k$-th analysis. As an example, set $(w_{1k}^{PFS}, w_{2k}^{PFS}) = \left(\sqrt{\frac{n_{1k,PFS}}{n_{1k,PFS}+n_{2k,PFS}}}, \sqrt{\frac{n_{2k,PFS}}{n_{1k,PFS}+n_{2k,PFS}}}\right)$ for PFS hypothesis where $n_{ik,PFS}$ is the number of PFS events from stage i subjects (i=1,2) and $(w_{1k}^{OS}, w_{2k}^{OS}) = \left(\sqrt{\frac{n_{1k,OS}}{n_{1k,OS}+n_{2k,OS}}}, \sqrt{\frac{n_{2k,OS}}{n_{1k,OS}+n_{2k,OS}}}\right)$ for OS hypothesis where $n_{ik,OS}$ is the number of OS events from stage $i$ subjects (i=1,2). However, $w_{1k}$ and $w_{2k}$ should be pre-specified in order to control the Type-I error rate. Since it is impossible to know the decision at futility analysis and the number of events from stage 1 and 2 for each efficacy analysis, we use pre-specified weights to compute p-values.

As mentioned in previous section, the proposed $g$GSD is FWER controlled and the simulations showed it is conservative; i.e., the type I error is less than the specified 0.025 level as shown in Table 3. Table 4 shows the power of rejecting the sub-group (S), or both sub-group and full population (S&F). The performance of the proposed $g$GSD depends on the choice of the weights $w_{1k}$ and $w_{2k}$. The first set of weights are computed using the number of PFS/OS events in the simulation and are used as a reference. When $w_{1k} < w_{2k}$, AD and $g$GSD have lower power to detect treatment efficacy compared with GSD. When $w_{1k} \geq w_{2k}$, $g$GSD has higher power than GSD and AD. Table 4 indicates that the events driven weight or more weights for stage 1 data lead to a better *g*GSD performance. The performance of *g*GSD is robust for the weights as long as more weight is assigned to stage 1 data. Thus, assigning more weights for data from stage 1 is recommended in order to utilize the information more efficiently. The simulation results for setting 3 where only sub-group has significant treatment benefit demonstrate that the proposed gGSD reduces the patient's exposure to less effective treatment comparing to GSD if the complementary sub-group has less significant treatment effect since gGSD will not enroll patients in the complementary sub-group in stage 2.

**Table 3:** Family-wise error rate for three simulation settings

| Weight($w_{1k}, w_{2k}$) | | Design | Setting 1 | Setting 2/3 |
|---|---|---|---|---|
| PFS | OS | GSD | 0.015 | 0.019 |
| $(w_{1k}^{PFS}, w_{2k}^{PFS})$ | $(w_{1k}^{OS}, w_{2k}^{OS})$ | AD | 0.010 | 0.009 |
| | | gGSD | 0.012 | 0.012 |
| $(\sqrt{0.2}, \sqrt{0.8})$ | $(\sqrt{0.2}, \sqrt{0.8})$ | AD | 0.008 | 0.008 |
| | | gGSD | 0.010 | 0.010 |
| $(\sqrt{0.3}, \sqrt{0.7})$ | $(\sqrt{0.3}, \sqrt{0.7})$ | AD | 0.009 | 0.008 |
| | | gGSD | 0.013 | 0.009 |
| $(\sqrt{0.5}, \sqrt{0.5})$ | $(\sqrt{0.5}, \sqrt{0.5})$ | AD | 0.011 | 0.010 |
| | | gGSD | 0.014 | 0.012 |
| $(\sqrt{0.5}, \sqrt{0.5})$ | $(\sqrt{0.7}, \sqrt{0.3})$ | AD | 0.011 | 0.010 |
| | | gGSD | 0.013 | 0.012 |
| $(\sqrt{0.7}, \sqrt{0.3})$ | $(\sqrt{0.7}, \sqrt{0.3})$ | AD | 0.010 | 0.009 |
| | | gGSD | 0.012 | 0.012 |
| $(\sqrt{0.8}, \sqrt{0.2})$ | $(\sqrt{0.8}, \sqrt{0.2})$ | AD | 0.010 | 0.009 |
| | | gGSD | 0.012 | 0.010 |
| $(\sqrt{0.6}, \sqrt{0.4})$ | $(\sqrt{0.6}, \sqrt{0.4})$ | AD | 0.011 | 0.010 |
| | | gGSD | 0.012 | 0.013 |

*: Weights defined in early text based on observed interim events and planned final events

**Table 4:** gGSD power for three simulation settings

| Weight($w_{1k}, w_{2k}$) | | Design | Setting 1 | | Setting 2 | | Setting 3 |
|---|---|---|---|---|---|---|---|
| PFS | OS | | S | S & F | S | S & F | S |
| $(w_{1k}^{PFS}, w_{2k}^{PFS})$* | $(w_{1k}^{OS}, w_{2k}^{OS})$* | GSD | 0.884 | 0.880 | 0.914 | 0.914 | 0.914 |
| | | AD | 0.910 | 0.920 | 0.948 | 0.952 | 0.891 |
| | | gGSD | 0.924 | 0.933 | 0.949 | 0.953 | 0.925 |
| $(\sqrt{0.2}, \sqrt{0.8})$ | $(\sqrt{0.2}, \sqrt{0.8})$ | AD | 0.740 | 0.741 | 0.834 | 0.835 | 0.793 |
| | | gGSD | 0.765 | 0.761 | 0.834 | 0.836 | 0.851 |
| $(\sqrt{0.3}, \sqrt{0.7})$ | $(\sqrt{0.3}, \sqrt{0.7})$ | AD | 0.804 | 0.806 | 0.879 | 0.881 | 0.840 |
| | | gGSD | 0.825 | 0.827 | 0.881 | 0.884 | 0.881 |
| $(\sqrt{0.5}, \sqrt{0.5})$ | $(\sqrt{0.5}, \sqrt{0.5})$ | AD | 0.872 | 0.880 | 0.932 | 0.936 | 0.885 |
| | | gGSD | 0.894 | 0.901 | 0.934 | 0.937 | 0.915 |
| $(\sqrt{0.5}, \sqrt{0.5})$ | $(\sqrt{0.7}, \sqrt{0.3})$ | AD | 0.896 | 0.905 | 0.946 | 0.949 | 0.890 |
| | | gGSD | 0.912 | 0.920 | 0.949 | 0.952 | 0.916 |
| $(\sqrt{0.7}, \sqrt{0.3})$ | $(\sqrt{0.7}, \sqrt{0.3})$ | AD | 0.910 | 0.919 | 0.948 | 0.952 | 0.892 |
| | | gGSD | 0.923 | 0.932 | 0.950 | 0.954 | 0.926 |
| $(\sqrt{0.8}, \sqrt{0.2})$ | $(\sqrt{0.8}, \sqrt{0.2})$ | AD | 0.909 | 0.918 | 0.948 | 0.952 | 0.883 |
| | | gGSD | 0.922 | 0.931 | 0.951 | 0.954 | 0.927 |
| $(\sqrt{0.6}, \sqrt{0.4})$ | $(\sqrt{0.6}, \sqrt{0.4})$ | AD | 0.894 | 0.903 | 0.943 | 0.947 | 0.892 |
| | | gGSD | 0.916 | 0.923 | 0.945 | 0.948 | 0.920 |

*: Weights defined in early text based on observed interim events and planned final events

Another advantage of the proposed gGSD is that it can terminate early given sufficient evidence. Figure 3 shows the termination time of three designs using the last three sets of weights in Table 4. For GSD, the trial is terminated if and only if all the four hypotheses become significant. For example, there are 3 hypotheses being significant in IA1 and the last hypothesis is rejected in IA2, then the termination time for this trial is at IA2. The circumstances for AD and gGSD are a little different since both designs include futility analysis and different groups may be selected for further investigation in stage 2. The trial stops at the futility stage if no group passes the futility analysis. A trial where at least one group passes the futility analysis is terminated if and only if all null hypotheses in the selected group are rejected. The trial continues if there is a null hypothesis in the selected group not yet being rejected. Figure 4 shows that the proposed gGSD stops early for all three scenarios. Among 2000 simulations, the percent of trials that continue to FA using gGSD is less than using GSD or AD. In other words, gGSD has higher probability to terminate early, suggesting that gGSD requires less time and resources to prove new treatment efficacy than GSD and AD without sacrificing power for an important underlying benefit.

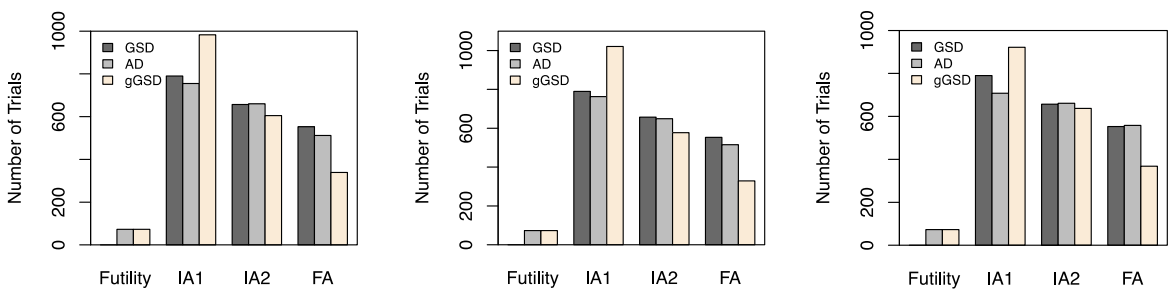

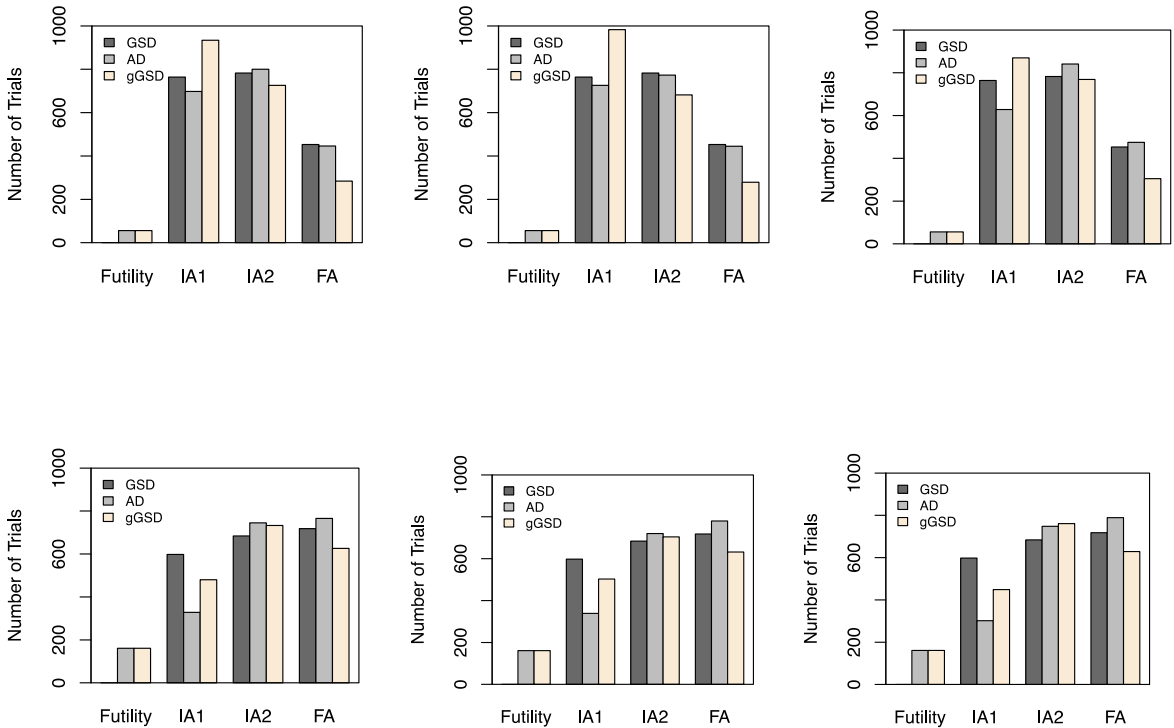

**Figure 3**: The time of efficiency analyses conducted among 2000 trials until all the 4 hypotheses are significant. Futility column denotes that trials terminate with no population group selected to continue past the futility analysis. The top 3 plots from left to right correspond to setting 1 with the last 3 sets of weights for inverse-normal combination test given in Table 4. The middle 3 plots and bottom 3 plots correspond to setting 2 and setting 3, respectively.

## 4. An illustrative example

We use an example with specified p-values to illustrate the potential advantage of the proposed *g*GSD compared to GSD. Consider a group sequential design for a Phase III 2$^{nd}$ line small cell lung cancer trial with a 50% prevalence of platinum-sensitive subgroup where PFS and OS are the dual primary endpoints. The graphical approach (Maurer and Bretz, 2013) was used to control FWER of the four hypotheses with a total of FWER level 0.025 in the GSD with the setting 2 parameters in Table 2 used. PFS and OS hypotheses are tested for the first two efficacy analyses and only OS hypotheses are tested at the final analysis. The nominal p-values at each interim analysis and the final analysis for both GSD and *g*GSD are shown in Table 5.

The data generated p-values at each interim analysis and the final analysis for the classical GSD are listed in Table 5. Table 5 indicates none of the four hypotheses are rejected by using GSD. Compared to GSD, *g*GSD increases the power to reject the null hypotheses. Using the *g*GSD and the gated rules in Table 1 with $\theta^F = 0.83$ and $\theta^S = 0.85$ at the end of stage 1, stage 2 is continued for the full group only with Scenario 2 described for stage 2 in Section 2, where $H_0^{\{F,OS\}}$ will be tested at level $\alpha = 0.025$ only if $H_0^{\{F,PFS\}}$ is rejected. Thus, no hypothesis is tested for the subgroup at all. A fixed weights $w_{1k}$ and $w_{2k}$ as $\sqrt{0.5}$ is used for all the p-value combination tests in $g$GSD. With a p-value of 0.0022 at the IA1 for the PFS, the PFS(F) is rejected. Thus, according to the rules for Scenario 2 for stage 2, OS(F) can be tested. With a p-value of 0.0125 at IA1, it is failed to reject OS(F) at IA1. Then the trial continues to IA2 for OS(F) testing only. With a p-value of 0.0019 at the IA2 for the OS, the OS(F) is rejected at IA2. Table 5 provides the detailed nominal p-value for each hypothesis at IAs and FA as well as the data generated p-values. None of the four hypotheses are rejected by using GSD while $g$GSD rejects two full group hypotheses. For $g$GSD, PFS(F) is rejected at IA1 and OS(F) is rejected at IA2.

**Table 5:** Theoretical and specified parameters for the illustrative example

| Analysis | Boundary & p-value | GSD ||||  gGSD  ||||
| | | SubGroup || Full Group || SubGroup || Full Group ||
| | | PFS | OS | PFS | OS | PFS | OS | PFS | OS |
|---|---|---|---|---|---|---|---|---|---|
| IA1 | Nominal p-value | 0.0036 | 0.0017 | <0.0001 | <0.0001 | 0.0037 | 0.0017 | 0.0039 | 0.0024 |
| | Data generated p-value | 0.0177 | 0.1205 | 0.0008 | 0.0104 | -- | -- | **0.0022** | 0.0125 |
| IA2 | Nominal p-value | 0.0088 | 0.0078 | 0.0002 | 0.0001 | 0.0090 | 0.0079 | 0.0089 | 0.0088 |
| | Data generated p-value | 0.0137 | 0.0502 | 0.00022 | 0.0023 | -- | -- | -- | **0.0019** |
| FA | Nominal p-value | | | 0.0120 | 0.0002 | | | 0.0122 | 0.0119 |
| | Data generated p-value | | | 0.0534 | 0.0011 | | | -- | -- |

Notes:
The nominal p-value is the p-value boundary in a typical group sequential design under the allocated alpha in different IA time.
The data generated p-value is the p-value from the test using the trial data.

## 5. Summary

Seamless Phase II/III designs are getting more attention and being increasingly adopted as a cost effective and time saving drug development strategy. In this paper, we proposed a gated group sequential design for seamless Phase II/III trial with potential sub-group selection. Combining this with GSD, our proposed $g$GSD design enables population selection and multiple interim analyses to enable early stopping. In this paper, we extended Jenkins, et al. (2011) method for population selection to control FWER for all four of the aforementioned hypotheses with dual primary endpoints. We further added a group sequential design strategy into stage 2 for flexible early efficacy testing. The hierarchical testing strategy proposed by Glimm et. al. (2010) was modified to accommodate our multiple testing scenarios with FWER controlled between populations, instead of the originally proposed usage for the ordered testing of endpoints such as PFS and OS with FWER. Within each population, the graphical approach combined with standard group sequential design was used for flexibility. The familywise error rate of proposed $g$GSD is strictly controlled. Sub-group and the full population are tested hierarchically to control the FWER. Simulation results and the illustrative example suggest that the gated group sequential design can increase power compared to the other trial designs; e.g., the proposed gated group sequential design could achieve the same power with a smaller sample size compared to the commonly used GSD. Furthermore, the trial can terminate early with sufficient strong evidence from efficacy analyses and potentially moves efficacious products into market faster for unmet medical needs. A special note on the particular advantage of the gGSD over GSD in the simulation study occurs when the true benefit is just in the sub-group. The gGSD has the ability to focus on the stage 2 selected population, increases power over a Phase 3 study of both populations and reduces the patient's exposure to less effective treatment comparing to GSD if the complementary sub-group has less significant treatment effect.

To design a clinical trial with power, say, 90%, using a gated group sequential design, the sample size can be calculated using a classical group sequential design with a non-binding stopping for futility with the same power 90%, or with a little smaller power, say, 85%, due to the fact that the $g$GSD has better power than the GSD with the same sample size. When using the sample size calculated from GSD using the same planned power 90%, the final power using $g$GSD in analyzing the trial data will be bigger than 90%. When analyzing the trial data using $g$GSD, more weight should be given as the pre-specified value for stage 1 data in the combination test to obtain better $g$GSD power as demonstrated in the simulations.

The idea proposed in this paper can also be applied to conduct efficient trials and simultaneously investigate several vital questions for drug development, such as identifying the most beneficial sub-group for a new treatment or dose (treatment) selection problem. Moreover, the proposed $g$GSD is applicable to more than one sub-group where the sub-groups are nested. In this paper, the sub-group was pre-specified. However, this sub-group information may not be always accurately identified before the trial. Freidlin and Simon (2005) proposed an adaptive signature design to find sensitive patients, without pre-specified, into a formal Phase III trial. The proposed seamless design shares the same potential operational challenges discussed in the literature that the trial team may choose to hold the enrollment while the team decides the population selection at the end of stage 1.

In this paper, the PFS of the dual-primary endpoints was used for the adaption. However, other surrogate "proof-of-concept" endpoint such as the objective response if that is more appropriate may be used. The $g$GSD is a two-stage trial design with two arms where the second stage data are used for a classical group sequential design framework. In this regard, the more recently commonly discussed multi-arm multi-stage (MAMS) design can be combined with $g$GSD. The research is under investigation.

**Declaration of conflicting interests**

The authors declared no potential conflicts of interest with respect to the research, authorship, and/or publication of this article.

**Funding**

This research received no specific grant from any funding agency in the public, commercial, or not-for-profit sectors.